\documentclass{article}
\usepackage{spconf,amsmath,graphicx,hyperref}
\usepackage{amssymb}
\usepackage{booktabs}
\usepackage{booktabs}
\usepackage{xcolor}
\usepackage{subcaption}
\usepackage{array}

\title{Scalable Evaluation for Audio Identification via Synthetic Latent Fingerprint Generation}

\name{Aditya Bhattacharjee, Marco Pasini, Emmanouil Benetos
\thanks{
A. Bhattacharjee and M. Pasini are research students at the UKRI Centre for Doctoral Training in Artificial Intelligence and Music, supported jointly by UK Research and Innovation [grant number EP/S022694/1] and Queen Mary University of London.}}

\address{School of Electronic Engineering and Computer Science,\\
Queen Mary University of London, UK}

\begin{document}
\ninept
\maketitle

\begin{abstract}
The evaluation of audio fingerprinting at a realistic scale is limited by the scarcity of large public music databases. We present an audio-free approach that synthesises latent fingerprints which approximate the distribution of real fingerprints. Our method trains a Rectified Flow model on embeddings extracted by pre-trained neural audio fingerprinting systems. The synthetic fingerprints generated using our system act as realistic distractors and enable the simulation of retrieval performance at a large scale without requiring additional audio. We assess the fidelity of synthetic fingerprints by comparing the distributions to real data. We further benchmark the retrieval performances across multiple state-of-the-art audio fingerprinting frameworks by augmenting real reference databases with synthetic distractors, and show that the scaling trends obtained with synthetic distractors closely track those obtained with real distractors. Finally, we scale the synthetic distractor database to model retrieval performance for very large databases, providing a practical metric of system scalability that does not depend on access to audio corpora.

\end{abstract}
\section{Introduction}
\label{sec:intro}

Audio fingerprinting refers to a set of techniques for identifying recordings by converting audio excerpts into compact embeddings that can be efficiently matched against a large reference database \cite{cano2005audio}. A fingerprint must be robust to distortions while remaining specific enough to uniquely identify the underlying recording. At query time, the excerpt is converted to a fingerprint and matched against the database of references. The performance of such systems depends on robustness, discriminability, and scalability to millions of entries under strict latency and memory budgets.

Recent approaches formulate audio fingerprinting as a self-supervised learning task, where neural encoders learn invariance to degradations such as noise, reverberation, or time-scale modification in a data-driven way. Training objectives typically use contrastive learning with strong augmentation, combined with efficient indexing for approximate nearest-neighbour retrieval. Encoder designs range from CNNs \cite{chang2021neural, araz2025enhancing} to transformer-based models \cite{singh2022attention, singh2023simultaneously, singh2024flowhash}, and more lightweight alternatives such as graph neural networks in \textit{GraFPrint} \cite{grafp_2025} or PointNet-inspired architectures in \textit{PeakNetFP} \cite{peaknetfp_2025}.

Scalability in audio fingerprinting refers to how retrieval performance degrades as the number of distractors grows. Realistic evaluation requires very large databases, since false matches and near-duplicates become more likely as the pool expands. However, the scarcity of public music datasets at scale due to copyright, licensing, and storage constraints means that state-of-the-art systems are rarely benchmarked at true real-world scale.

To address this, we propose a generative modelling approach that synthesises realistic fingerprints directly in latent space, without requiring audio. We use a Rectified Flow model \cite{liu2023flow}, a diffusion-based method that learns a velocity field transforming Gaussian noise into fingerprint embeddings from the same distribution as those produced by pre-trained systems. This enables high-fidelity sampling of synthetic fingerprints that serve as realistic distractors, making it possible to simulate large-scale retrieval scenarios using only embeddings.

These are our main contributions:
\begin{enumerate}
    \item We introduce a Rectified Flow model for generating synthetic distractors that approximate the distribution of real audio fingerprints.
    \item We compare real and synthetic fingerprint distributions, and evaluate synthetic samples as distractors in audio identification tasks.
    \item We assess the scalability of state-of-the-art fingerprinting systems using large-scale synthetic distractors as a proxy for million-song databases.
\end{enumerate}

All code and trained models are open-sourced at: \url{https://github.com/chymaera96/audio-id-at-scale}.


\begin{figure}[t]
\centering
\includegraphics[width=0.5\textwidth, trim={0cm 1.5cm 0cm 1.5cm}, clip]{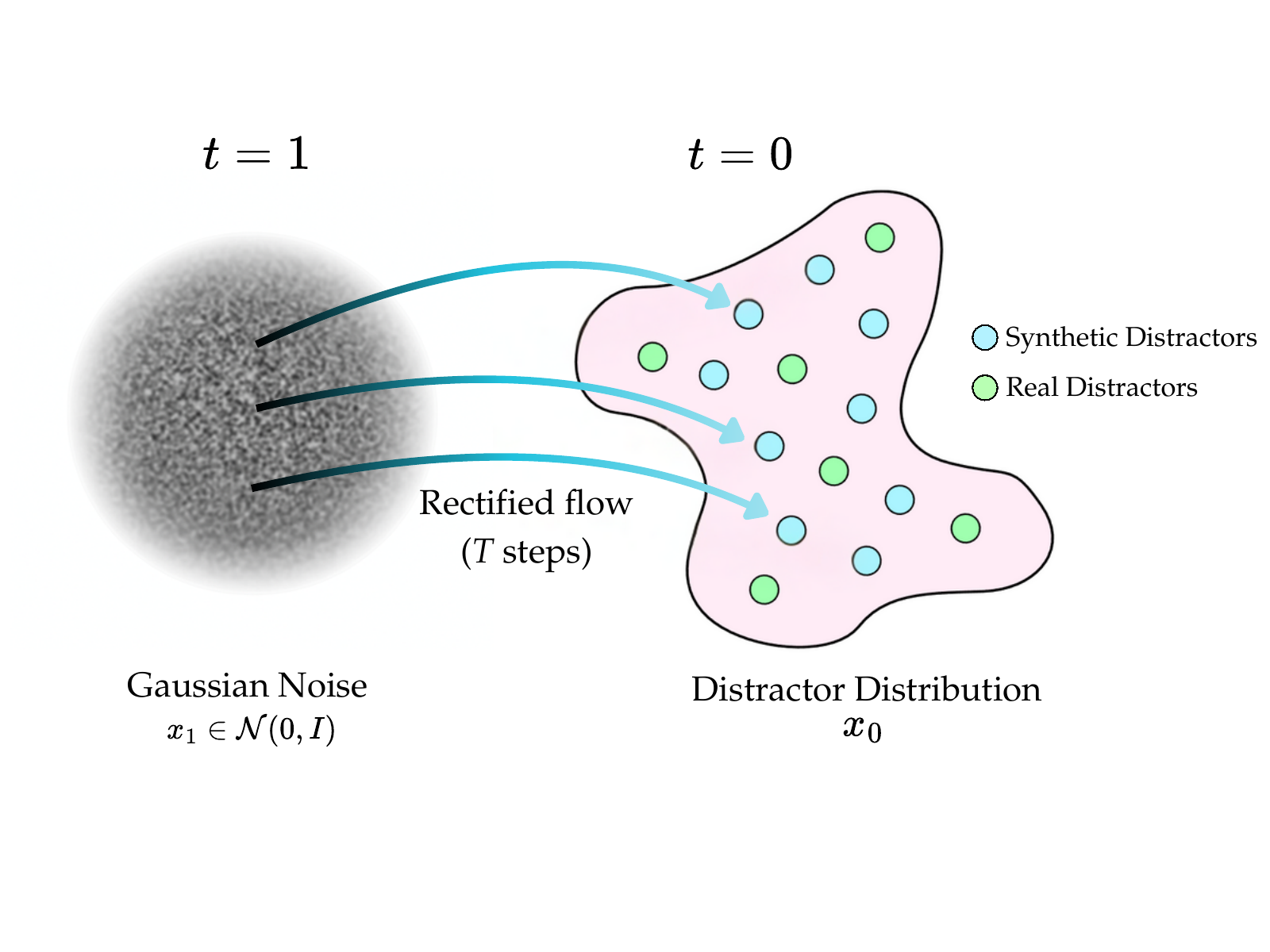}
\caption{A Rectified Flow model is used to generate distractors from noise samples. The generated distractors are then used to populate the distractor set with an arbitrary density.}
\label{fig:model}
\end{figure}

\section{Rectified Flow Model}
\label{sec:rf-model}

Our goal is to learn the latent distribution of fingerprint embeddings produced by a fixed, pre-trained neural audio fingerprinting system. These embeddings capture semantic and perceptual characteristics of audio clips, and the resulting latent space varies across different fingerprinting models. Therefore, we train a separate generative model for four open-sourced audio fingerprinting systems (refer to \ref{sec:baselines}, without assuming a unified embedding space. We adopt the Rectified Flow framework, which enables us to synthesise new fingerprint-like embeddings that lie on the manifold of real data.

\subsection{Problem Formulation}

Let \( \mathcal{F}: \text{audio} \rightarrow \mathbb{R}^d \) be a fixed, pre-trained fingerprinting model that maps 1-second audio excerpts to latent embeddings in \( \mathbb{R}^d \). Each audio signal is divided into overlapping windows of 1 second with a 50\% overlap, and passed through \( \mathcal{F} \) to yield a sequence of fingerprint vectors.

We denote a fingerprint embedding by \(x \in \mathbb{R}^d\). Given a database \( \mathcal{D}_n \) consisting of \( n \) songs, the empirical distribution of all fingerprints extracted from \( \mathcal{D}_n \) is denoted by \( p_n(x) \). Our objective is to model the fingerprint distribution \( p_{n'}(x) \) associated with a much larger, hypothetical database \( \mathcal{D}_{n'} \), where \( n' \gg n \).

To approximate this distribution, we train a generative model \( \mathcal{G}_\theta: z \mapsto \tilde{x} \), where \( z \sim \mathcal{N}(0, I) \) and \( \tilde{x} \in \mathbb{R}^d \) represents a synthetic fingerprint sample. The model \( \mathcal{G}_\theta \) learns to transform noise vectors into samples from the data manifold defined by \( p_n(x) \), thereby enabling us to simulate retrieval at larger database scales.

\subsection{Model Architecture}

The model is implemented as a multi-layer perceptron (MLP) conditioned on a scalar diffusion time variable \( t \in [0, 1] \). The conditioning is applied via sinusoidal time embeddings \cite{vaswani2017attention} denoted as \( \tau(t) \in \mathbb{R}^{d_\tau} \), where \( d_\tau \) is the embedding dimension. At each layer, the MLP processes the current input while incorporating \( \tau(t) \) to modulate the transformation via Adaptive Layer Normalisation (AdaLN).\footnote{The diffusion time variable \(t\) parameterises the interpolation between data and noise in the rectified flow model. It should not be confused with the temporal indexing of fingerprints within the audio signal.}

The architecture is summarised in Table~\ref{tab:rf-architecture}.

\begin{table}[h]
\centering
\caption{Network architecture for Rectified Flow model}
\label{tab:rf-architecture}
\begin{tabular}{@{}lll@{}}
\toprule
\textbf{Component} & \textbf{Configuration} & \textbf{Dimension/Width} \\
\midrule
Input & Latent vector sample & $d=128$ \\
Time embedding & Sinusoidal & $d_\tau = 32$\\
Input projection & Linear layer & $d$ \( \rightarrow \) 768 \\
Hidden layers & 12 MLP blocks & 768 \( \rightarrow \) 3072 \( \rightarrow \) 768 \\
Normalization & Adaptive layer norm & - \\
Output layer & Linear layer & 768 \( \rightarrow \) $d$ \\
\bottomrule
\end{tabular}
\end{table}

\subsection{Training}

We define a forward process that interpolates between a real fingerprint \( x \in \mathbb{R}^d \) and Gaussian noise \( z \sim \mathcal{N}(0, I) \) using a diffusion time variable \( t \in [0, 1] \):
\begin{equation}
x_t = t \cdot z + (1 - t) \cdot x.
\end{equation}
Here, \(t=0\) corresponds to the clean fingerprint, and \(t=1\) corresponds to pure noise. The rectified flow objective is to predict the velocity field that transforms data towards noise:
\begin{equation}
v(x_t, t) = z - x.
\end{equation}
The network is trained to approximate this velocity by minimising the mean squared error:
\begin{equation}
\mathcal{L}(\theta) = \mathbb{E}_{x, z, t} \left[ \left\| \hat{v}_\theta(x_t, t) - (z - x) \right\|^2 \right].
\end{equation}
Training is performed on normalised fingerprint embeddings using the empirical mean and variance of the dataset. The model does not require class labels, augmentations, or audio input.

\subsection{Sampling Procedure}

To generate synthetic fingerprints, we initialise with a noise vector \( x_1 \sim \mathcal{N}(0, I) \) and integrate the learned velocity field backwards from \(t=1\) to \(t=0\). Using an explicit Euler scheme with \(T\) steps and \(\Delta t = 1/T\), the update rule is:
\begin{equation}
x_{t - \Delta t} = x_t + \Delta t \cdot \hat{v}_\theta(x_t, t).
\end{equation}
This process is repeated until \(x_0\) is reached, corresponding to a generated fingerprint sample. Finally, the sample is rescaled using the training set statistics. The resulting vectors follow the learned latent distribution and can be used as distractors during large-scale retrieval evaluation. The sampling procedure has been illustrated in Figure \ref{fig:model}.

\subsection{Implementation}
\label{sec:implementation}
We trained the rectified flow model for 100 epochs on a single NVIDIA A100 GPU. The training process is lightweight, requiring roughly one hour to converge. Optimisation was performed using AdamW \cite{loshchilov2017adamw} with a cosine decay learning rate schedule ranging from $5 \times 10^{-5}$ to $1 \times 10^{-6}$. Model validation was based on the Fréchet Distance \cite{heusel2017gans} between the generated fingerprints and the training set, providing a distributional measure of generation fidelity.

\section{Dataset}
\label{sec:dataset}

We use the Free Music Archive (FMA) dataset \cite{defferrard2016fma} to construct both the training and evaluation databases. The generative model is trained on fingerprint embeddings extracted from the FMA medium subset, which contains 25,000 songs across a range of genres.

For evaluation, we use tracks from the FMA large subset to build two disjoint sets: a \textit{query}-\textit{reference} database and a \textit{distractor} database. The reference database contains clean fingerprint embeddings used as the retrieval target. The query set consists of 500 embeddings derived from 1-second excerpts of reference audio, each subjected to common distortions as specified in the benchmarked fingerprinting models (see Section~\ref{sec:baselines}). Real distractors are constructed from unrelated tracks in FMA large and used to simulate large-scale retrieval scenarios.

\section{Evaluation}
\label{sec:evaluation}

The primary objective of our evaluation methodology is to assess whether synthetic distractors generated by our model can serve as viable substitutes for real distractors in stress-testing audio fingerprinting systems. In order to evaluate the effectiveness of our approach, we focus on two axes of evaluation: (i) the fidelity of the synthetic fingerprints compared to real fingerprints in latent space, and (ii) the impact of synthetic distractors on retrieval performance across a range of database sizes. 

\begin{figure*}[t]
    \centering
    \includegraphics[width=0.8\linewidth]{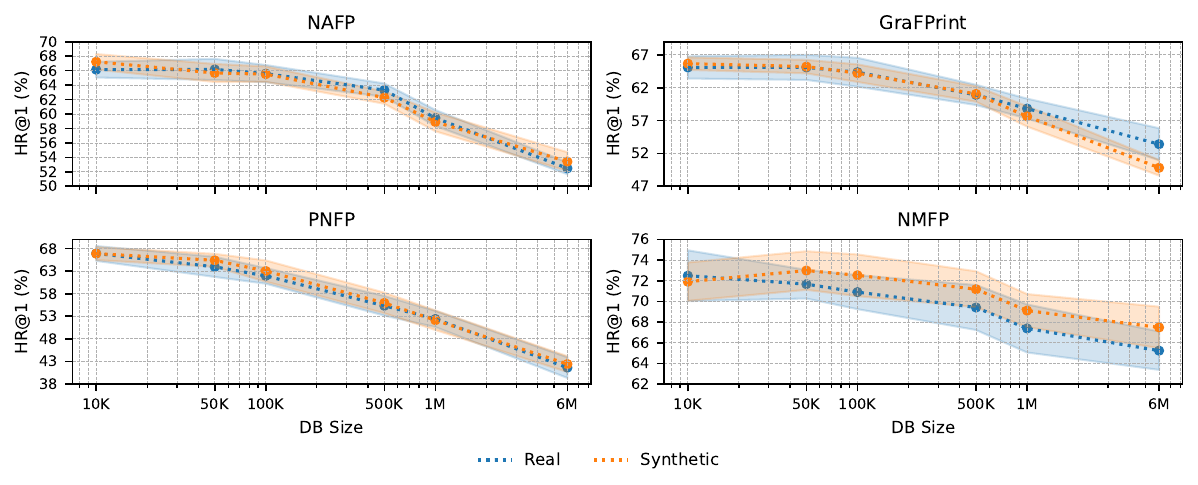}
    \caption{Top-1 hit rate as a function of distractor database size for four fingerprinting systems. Synthetic distractors closely track real-data scaling trends across all models. The shaded region represents $\pm1$ standard deviation, and the dots show the mean hit rate.}
    \label{fig:scaling-curves}
\end{figure*}

\subsection{Benchmarked Fingerprinting Models}
\label{sec:baselines}

We benchmark several open-sourced audio fingerprinting systems using our evaluation framework. Each model employs a distinct training-time augmentation strategy to enforce invariance to perturbations such as noise mixing, time-stretching, or reverberation. Table~\ref{tab:fp-models} summarizes their encoder architectures and augmentation schemes. For fairness, we construct a query set for each model using its native perturbation recipe. Rather than comparing models directly, which would conflate their differing invariance objectives, we report retrieval performance under scaling with real versus synthetic distractors for each model independently. To enable efficient large-scale benchmarking, we employ IVF-PQ \cite{johnson2019billion} for approximate nearest-neighbour (ANN) search, consistent with the native retrieval setups used in these works.

\begin{table}[h]
\centering
\caption{Audio fingerprinting models benchmarked in this work}
\label{tab:fp-models}
\begin{tabular}{@{}llp{4.2cm}@{}}
\toprule
\textbf{Framework} & \textbf{Encoder} & \textbf{Perturbation Strategy} \\
\midrule
NAFP \cite{chang2021neural} & CNN &
Background noise \newline
Short IR reverb ($\leq$75\,ms) \\
\midrule
GraFPrint \cite{grafp_2025} & GNN &
Background noise \newline
Full IR reverb \\
\midrule
PeakNetFP \cite{peaknetfp_2025} & PointNet++ &
Time-stretching \\
\midrule
NMFP \cite{araz2025enhancing} & CNN &
Background noise \newline
Full IR reverb \\
\bottomrule
\end{tabular}
\end{table}

\subsection{Synthetic Fingerprint Fidelity}
\label{sec:fidelity}

We evaluate the fidelity of synthetic fingerprints by comparing their distribution to that of real embeddings, and contrasting both against Gaussian noise as a baseline. Two complementary metrics are employed: (i) the Fréchet Distance (FD), computed between the mean and covariance of real and synthetic sets, and (ii) the Jensen–Shannon (JS) divergence. For the latter, distributions are estimated with Gaussian kernel density estimation (KDE) after projecting embeddings onto the top 20 principal components of the real dataset. Table~\ref{tab:scaling-fidelity} reports results for both comparisons. Synthetic fingerprints align closely with the real distribution, yielding low FD and JS scores, while their divergence from Gaussian noise is substantially larger. This indicates that the generative model can capture the statistical structure of fingerprints rather than producing unstructured vectors.

\begin{figure}[t]
\centering
\begin{subfigure}{0.48\linewidth}
    \includegraphics[width=\linewidth]{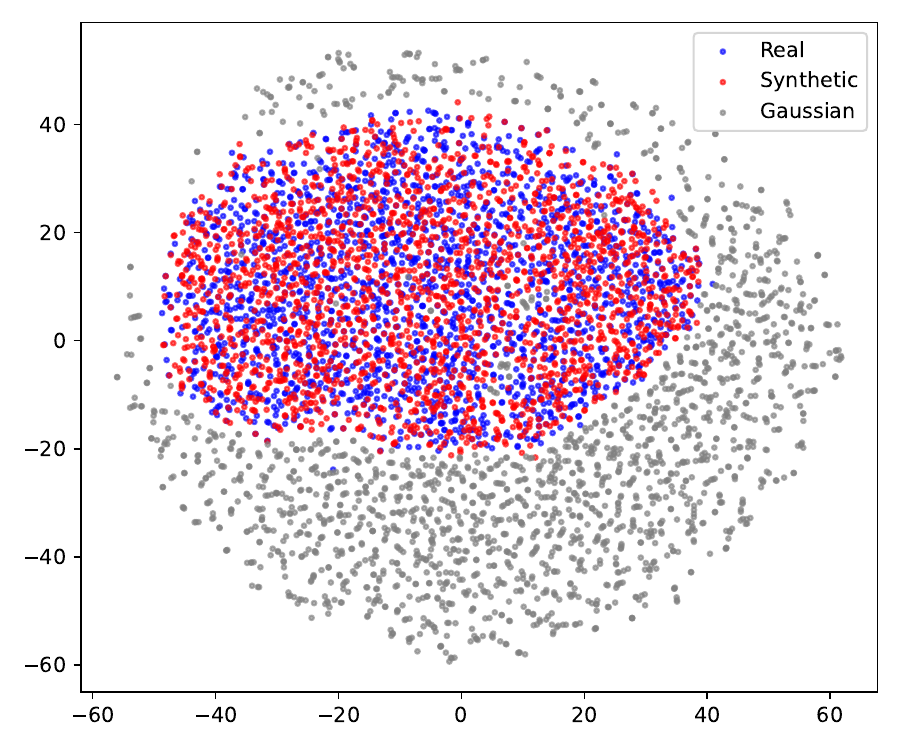}
    \caption{NAFP}
\end{subfigure}
\hfill
\begin{subfigure}{0.48\linewidth}
    \includegraphics[width=\linewidth]{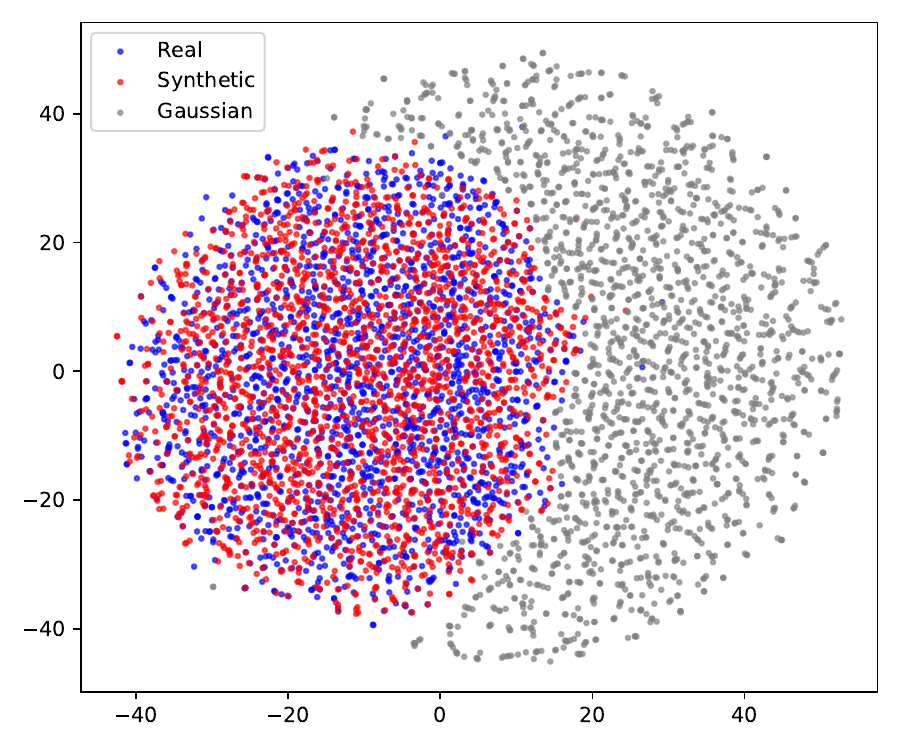}
    \caption{GraFPrint}
\end{subfigure}

\vspace{0.3em}

\begin{subfigure}{0.48\linewidth}
    \includegraphics[width=\linewidth]{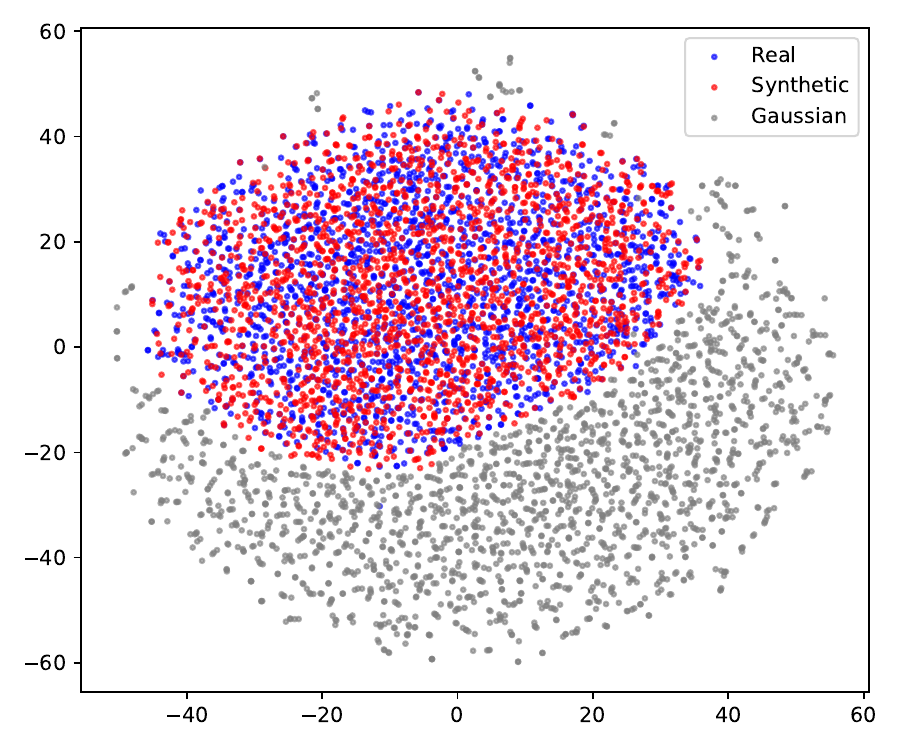}
    \caption{PeakNetFP}
\end{subfigure}
\hfill
\begin{subfigure}{0.48\linewidth}
    \includegraphics[width=\linewidth]{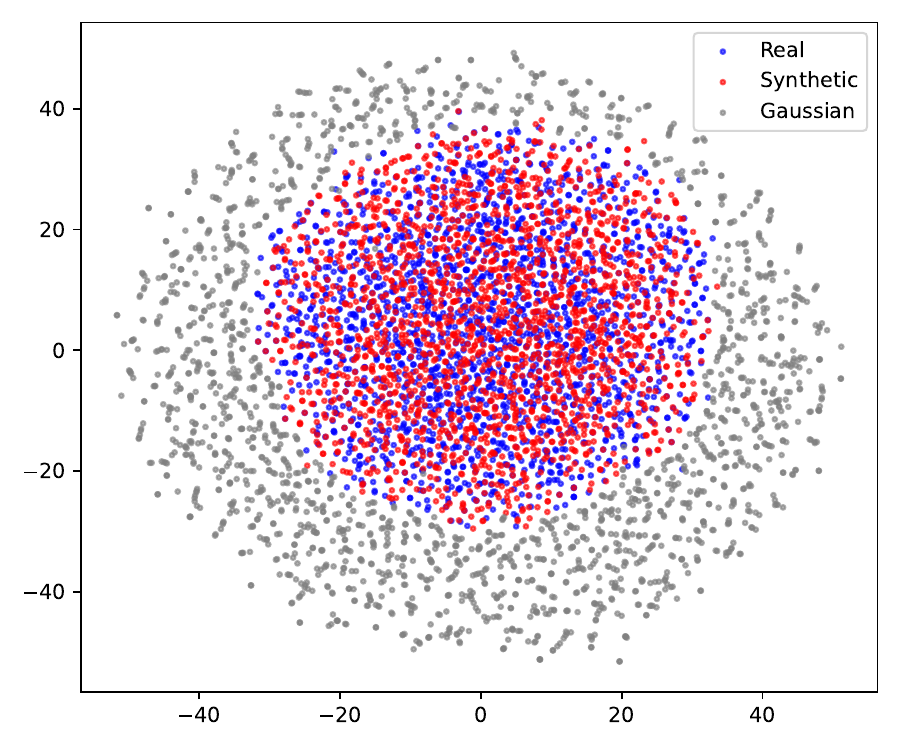}
    \caption{NMFP}
\end{subfigure}
\caption{t-SNE plots showing synthetic and real fingerprint distribution in comparison to Gaussian noise.}
\label{fig:tsne_grid}
\end{figure}

As a qualitative example, Figure~\ref{fig:tsne_grid} shows a t-SNE projection of the embedding space, where synthetic points are distributed consistently within the real manifold.

\subsection{Retrieval Performance under Scaling}
\label{sec:retrieval-scaling}

To evaluate scalability, we measure the top-1 hit rate for correct match retrieval as a function of database size. For each fingerprinting model, a fixed reference set and model-specific query set are constructed, where each query is derived by applying the model’s native perturbation strategy to a 1-second segment from the reference set (see Section~\ref{sec:baselines}).

The database is then augmented with varying numbers of non-matching distractors. These are either drawn from a pool of real fingerprints extracted from the \textit{fma-large} subset or generated using the rectified flow model. To account for sampling variability, we repeat each evaluation with multiple independently resampled distractor sets, both real and synthetic, and report the mean and standard deviation of the top-1 hit rate.

The top-1 hit rate, is defined as:
\begin{equation}
\text{HR@1} = \frac{1}{N} \sum_{i=1}^{N} \mathbb{1}\{y_i = \hat{y}_i^{(1)}\}
\end{equation}
where \( N \) is the number of queries, \( y_i \) is the ground truth match for query \( i \), and \( \hat{y}_i^{(1)} \) is the top-ranked result returned by the retrieval system.

This setup yields a retrieval degradation curve that reflects how performance deteriorates as the search space grows. Figure~\ref{fig:scaling-curves} compares these curves for real and synthetic distractors across all benchmarked models. We observe that the mean retrieval rates for real and synthetic distractors correspond closely. The overlapping shaded region indicates that the variance across repeated trials largely covers both conditions, suggesting that synthetic distractors impose a retrieval burden comparable to real distractors.

In all cases, the hit rate declines significantly with increasing distractors, highlighting that scalability remains a fundamental limitation across fingerprinting models, independent of their learned invariances.

\subsection{Extrapolating Retrieval at Scale}
\label{sec:scaling-extrapolation}

To compare the scalability of fingerprinting systems beyond available datasets, we simulate retrieval performance for a database containing 100 million fingerprints. Since no real dataset of this size is accessible, synthetic distractors sampled from the trained rectified flow model allow us to project the performance of each model in large-scale settings.

We define the percentage degradation in top-1 hit rate relative to the 1M scale as:
\begin{equation}
\text{Degradation}(\%) = 100 \times \frac{\text{HR@1}_{\text{1M}} - \text{HR@1}_{\text{100M}}}{\text{HR@1}_{\text{1M}}}
\end{equation}

\begin{table}[h]
\centering
\caption{Retrieval performance at scale and fidelity metrics. HR@1 reported at 1M and 100M synthetic distractors. JS: Jensen--Shannon divergence $D_{\mathrm{JS}}$, , FD: Fréchet Distance $D_{\mathrm{FD}}$. Row labels use $R=$ Real, $S=$ Synthetic, $G=$ Gaussian}
\label{tab:scaling-fidelity}
\begin{tabular}{@{}lcccc@{}}
\toprule
 & \textbf{NAFP} & \textbf{GraFPrint} & \textbf{PeakNetFP} & \textbf{NMFP} \\
\midrule
HR@1 \% (1M)        & 59.45 & 57.64 & 52.12 & 69.11 \\
HR@1 \% (100M)      & 37.77 & 39.65 & 23.32 & 59.26 \\
Degradation \% & 36.47 & 31.21 & 55.26 & \textbf{14.25} \\
\midrule
$D_{\mathrm{JS}}(\text{S}\,\|\,\text{G})$ & 0.676 & 0.693 & 0.678 & 0.6931 \\
$D_{\mathrm{JS}}(\text{S}\,\|\,\text{R})$  & 0.004 & 0.011 & 0.006 & 0.002 \\
$D_{\mathrm{FD}}(\text{S},\text{R})$  & 3.1e-3 & 2.3e-3 & 4.6e-3 & 6.5e-3\\

\bottomrule
\end{tabular}
\end{table}

Table~\ref{tab:scaling-fidelity} shows the top-1 hit rates for each model at 1M and 100M distractors, along with the computed degradation percentages. While all models show some degree of performance drop, the severity varies, highlighting differing levels of scalability among architectures. \textit{PeakNetFP} is a lightweight model which is robust to time-stretching to a very large degree. However, it exhibits a degradation of 55.26\% and may not be a viable framework at scale. While \textit{GraFPrint} and \textit{NMFP} are trained with similar perturbation strategies, the latter outperforms the scalability tests with considerably lower degradation in retrieval performance.

\section{Discussion}
\label{sec:discussion}

A key advantage of our approach is that it enables the simulation of large-scale retrieval scenarios without access to real audio. Extrapolating to 100M-scale databases with real data would require computing and storing billions of fingerprint vectors, which is prohibitively expensive in both latency and memory. By generating fingerprints directly in latent space, our method offers an efficient and reproducible way to assess scalability, effectively serving as a drop-in evaluation tool for fingerprinting systems when large annotated corpora are unavailable. Under the setup in Section~\ref{sec:implementation}, a synthetic database of 6 million fingerprints (corresponding to the database size for 100,000 30-second-long songs) can be produced in 25 minutes, whereas computing real fingerprints of the same size would take up to 10 hours, even with optimised parallelisation.

While our evaluation uses IVF-PQ for approximate nearest neighbour (ANN) search, similar degradation trends are observed with alternative backends such as locality-sensitive hashing (LSH)~\cite{indyk1998approximate} and deep hashing methods~\cite{singh2024flowhash, singh2023simultaneously}. This suggests that retrieval decline under scale is not tied to a specific indexing scheme, but may stem from broader modelling limitations or embedding overlap.

A potential direction for improving scalability is to structure the reference database into subsets that can be queried more selectively. For example, one could partition the database based on semantic similarity, artist identity, or expected query frequency derived from usage data. During retrieval, the system could prioritise higher-probability subsets first, either in parallel or sequentially, reducing the effective search space while maintaining accuracy. Our framework provides a practical platform to study such strategies by modelling the scaling behaviour of fingerprinting systems without reliance on massive annotated corpora.

\section{Conclusion}

We introduced a framework for scalable evaluation of audio fingerprinting systems using synthetic latent fingerprints generated by a rectified flow model. This approach addresses the practical challenge of benchmarking identification at million-scale when large annotated audio corpora are unavailable. By closely matching the distribution of real embeddings, synthetic fingerprints provide reliable proxies for real distractors, enabling systematic study of robustness and failure modes under realistic database growth. As future work, we will explore extending this framework to deep hashing methods for audio fingerprinting, as well as investigating database organisation strategies (e.g., semantic partitioning) to mitigate scalability bottlenecks in large-scale retrieval.

\vfill\pagebreak

\bibliographystyle{IEEEbib}
\bibliography{refs}

\end{document}